\documentclass[twocolumn,showpacs,amsmath,amssymb,pra,floatfix,aps,10pt]{revtex4-1}
\usepackage{graphicx}
\usepackage{dcolumn}
\usepackage{bm}
\begin{document}
%
\title{Energy quantization  at the three-quarter Dirac point in a magnetic field}
\author{Yasumasa Hasegawa and Keita Kishigi$^1$}
\affiliation
{Graduate School of Material Science, 
University of Hyogo, \\
3-2-1 Kouto, Kamigori, Hyogo, 678-1297, Japan, \\
$^1$ Faculty of Education, Kumamoto University, Kurokami 2-40-1, 
Kumamoto, 860-8555, Japan
}
\date{\today}

\begin{abstract}
The quantization of the energy in a magnetic field (Landau quantization) 
at a three-quarter Dirac point is studied theoretically. 
The three-quarter Dirac point is realized in the system of massless Dirac fermions 
with the critically tilted Dirac cone in one direction, where a linear term disappears and a 
quadratic term $\alpha_2 q_x^2$ with a constant $\alpha_2$ plays an important role. 
The energy is obtained as $E_n \propto \alpha_2^{\frac{3}{5}} (n B)^{\frac{4}{5}}$,
where $n=1, 2, 3, \dots$,
by means of numerically solving the differential equation. The same result is obtained    
 analytically by adopting an approximation.
The result is consistent with the semiclassical quantization rule studied previously. 
The existence of the $n=0$ state is studied by 
 introducing the energy gap due to the inversion-symmetry-breaking term, and it is obtained 
 that the $n=0$ state exists
 in one of a pair of three-quarter Dirac points, depending on the direction of
the magnetic field when the energy gap is finite. 
\end{abstract}

\maketitle

\section{Introduction}
Massless Dirac fermions are observed  in condensed matter physics,
in graphene \cite{Novoselov2005,Zhang2005}, 
organic conductors\cite{Tajima2000,Katayama2006,Kajita2014},
and the surface of the 3D topological insulators\cite{Fu2007,Hsieh2009}.  

When a two-dimensional system has an inversion symmetry and a time reversal symmetry, 
massless Dirac points ($\pm \mathbf{k}_{\textrm{D}}$) appear as a pair.
The minimal model for the massless Dirac fermions 
is written as\cite{Kobayashi2007,Goerbig2008,Kobayashi2009}
\begin{equation}
 \mathcal{H}_{\mathrm{D}} = 
 \begin{pmatrix}
   w_{0x} q_x +w_{0y} q_y & w_x q_x  \mp i w_y q_y\\
 w_x q_x \pm i w_y q_y &  w_{0x} q_x  +w_{0y} q_y 
\end{pmatrix},
\label{eqeq1}
\end{equation}
where 
\begin{equation}
\mathbf{q} = \mathbf{k} \mp \mathbf{k}_{\textrm{D}}.
\end{equation}
%
\begin{figure}[b]
\hspace{-0.5cm}
\centering
\includegraphics[width=0.4\textwidth]{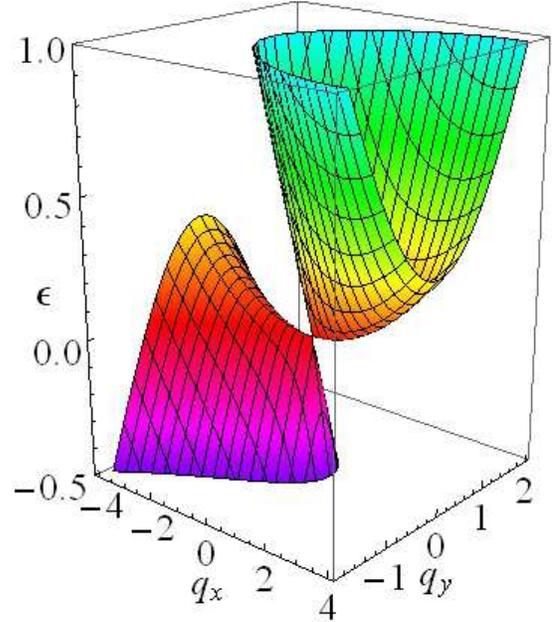} \\ 
\caption{(Color online) 
Energy as a function of $q_x$ and $q_y$ at $B=0$ at the three-quarter Dirac point.
Parameters are $w_x = 0.4$, $w_y=1$, $w_{0x}=-w_x$, $w_{0y}=0$,
$\alpha_2'=\alpha_2''=0.01$,
}
\label{figbandtqD}
\end{figure}
%
Two bands touch at the Dirac points. When
$w_{0x} = 0$ and $w_{0y} = 0$, the linear energy dispersion near the Dirac point (Dirac cone) is not tilted. 
By the finite $w_{0x}$ or $w_{0y}$, the Dirac cone is tilted, and
if the condition
\begin{equation}
\left(\frac{w_{0x}} {w_x}\right)^2 + \left(\frac{w_{0y}} {w_y}\right)^2 = 1,
\end{equation}
is fulfilled, the Dirac cone is critically tilted, i.e.,
the conical edge of the Dirac cone is horizontal in one direction.
In that case we have to take into account 
the quadratic terms in the tilted direction, except for the special case that the quadratic terms
vanish by symmetry or by accident.
Generally the quadratic terms exist as we have found previously\cite{Kishigi2017}
 in the tight-binding model
with pressure-dependent hoppings for the organic conductor,
 $\alpha$-(BEDT-TTF)$_2$I$_3$.
 The energy near the critically tilted Dirac point is shown in Fig.~\ref{figbandtqD}.
Since the energy of the upper band depends linearly in three directions 
(for example, $-q_x$ and $\pm q_y$)
and quadratically in one direction (for example, $+ q_x$) in that case, 
 we call the critically tilted Dirac point as the ``three-quarter Dirac point''.\cite{Kishigi2017}
 It has been known that when two-Dirac points merge at the time-reversal-invariant momentum,
 the energy depends linearly in two directions and quadratically in two directions, 
 and it is called the semi-Dirac 
point\cite{Hasegawa2006,Dietl2008,Montambaux2009,Banerjee2009,Delplace2010}.

 Previously we have shown that the energy in a magnetic field
(the Landau level) at the three-quarter Dirac point depends on the quantum number $n$ and 
the magnetic field $B$ as 
 \begin{equation}
 \epsilon_n \propto (n B)^{\frac{4}{5}} ,
\end{equation}
by calculating the energy of the tight-binding model for $\alpha$-(BEDT-TTF)$_2$I$_3$
in a magnetic field numerically\cite{Kishigi2017}. 
In that paper we have 
explained these $n$ and $B$ dependences of Landau levels by using
the semi-classical quantization rule. 
In this paper we 
study the Landau quantization
at the three-quarter Dirac point 
in a numerical study and an analytical treatment with a crude approximation.
The Dirac cone is taken to be critically tilted in the $k_x$ direction, i.e., $w_{0x} = -w_x$, and $w_{0y}=0$
in Eq.~(\ref{eqeq1}).
For simplicity we take $w_{x}>0$, $w_{y}>0$, 
and we introduce the quadratic terms in the $q_x$ direction 
($\alpha_2' q_x^2$ in diagonal elements and 
 $\alpha_2'' q_x^2$ in off-diagonal elements). 
Then the three-quarter Dirac Hamiltonian we study in this paper is
 \begin{equation}
 \mathcal{H}_{\mathrm{tqD}} = 
 \begin{pmatrix}
   - w_x q_x + \alpha_2' q_x^2 &  w_x q_x  +\alpha_2'' q_x^2 - i w_y  q_y\\
  w_x q_x +\alpha_2'' q_x^2 + i w_y q_y &  - w_x q_x + \alpha_2' q_x^2 
\end{pmatrix}.
\label{eqtqd0}
\end{equation}
\section{three-quarter Dirac point}
\subsection{energy at $B=0$}
In the absence of the magnetic field the energy is obtained by
\begin{equation}
 \mathcal{H}_{\mathrm{tqD}}  \Psi =  E (\mathbf{q}) \Psi ,
 \label{eqeq0}
\end{equation}
where
$\Psi$ is a wave function which has two components, $\psi_1$ and $\Psi_2$.
The eigenvalues of $\mathcal{H}_{\mathrm{tqD}}$ is obtained as
$E(\mathbf{q})=\varepsilon^0_{\mathrm{tqD}_{\pm}} (\mathbf{q})$;
 \begin{align}
  \varepsilon^0_{\textrm{tqD}_{\pm}} (\mathbf{q}) =& -w_x q_x + \alpha_{2}' q_x^2 \nonumber \\
  & \pm \sqrt{(w_x q_x+\alpha_{2}'' q_x^2)^2 + (w_y q_y)^2},
\end{align}
which are plotted in Fig.~\ref{figbandtqD}.
There exist the upper band  ($\varepsilon^0_{\textrm{tqD}_+}(\mathbf{q})$) 
and the lower band ($\varepsilon^0_{\textrm{tqD}_-}(\mathbf{q})$).
These two bands touch at $\mathbf{q}=(0,0)$.
Along the $q_x$ axis, the linear term disappears in $\varepsilon^0_{\textrm{tqD}_{+}} (\mathbf{q})$
and $\varepsilon^0_{\textrm{tqD}_{-}} (\mathbf{q})$ 
for $q_x >0$ and $q_x <0$, respectively, whereas
in other three directions the linear term exists;
\begin{align}
 \varepsilon_{\textrm{tqD}_+} (q_x, q_y=0) &= 
 \left\{ \begin{array}{ll}
 \alpha_2 q_x^2              & \mbox{ if $q_x>0$} \\
 2 w_x |q_x| + \tilde{\alpha}_2 q_x^2  & \mbox{ if $q_x < 0$}
\end{array} \right.  \label{eq07} \\
 \varepsilon_{\textrm{tqD}_+} (q_x=0, q_y) &= w_y |q_y| \\
 \varepsilon_{\textrm{tqD}_-} (q_x, q_y=0) &= 
 \left\{ \begin{array}{ll}
-2w_x q_x + \tilde{\alpha}_2 q_x^2  & \mbox{ if $q_x > 0$} \\
   \alpha_2 q_x^2              & \mbox{ if $q_x < 0$}
\end{array} \right. \label{eq09} \\
 \varepsilon_{\textrm{tqD}_-} (q_x=0, q_y) &= - w_y |q_y| ,
\end{align}
where 
\begin{equation}
  \alpha_2 = \alpha_2' + |\alpha_2''| , 
  \label{eqalpha2}
\end{equation}
and
\begin{equation}
  \tilde{\alpha}_2 = \alpha_2' - |\alpha_2''| .
\label{eqtildealpha2}
\end{equation}
If $\alpha_2>0$, $\mathbf{q}=0$ is a local minimum of $\varepsilon_{\textrm{tqD}_+}$
with the linear dispersion in three directions ($q_x<0$, $q_y>0$ and $q_y <0$) 
 and quadratic dispersion in one direction ($q_x>0$).
 Note that the three-quarter Dirac point
is neither the local maximum nor the local minimum of $\varepsilon_{\textrm{tqD}_-}$ if $\alpha_2>0$.
If $\alpha_2<0$, the  three-quarter Dirac point is the local maximum of   $\varepsilon_{\textrm{tqD}_-}$,
but it is  neither the local maximum nor the local minimum of $\varepsilon_{\textrm{tqD}_+}$.

\subsection{numerical results of the  energy at $B > 0$, using boundary condition at $y>0$}
\label{sectionbc1}
Hereafter we study the case $\alpha_2 >0$, i.e., the three-quarter Dirac point is the minimum of 
$\varepsilon^0_{\mathrm{tqD}_+}$, as shown in Fig.~\ref{figbandtqD}. 
In this case it is expected that when the magnetic field is applied,
there are the almost-localized bound states
(the Landau levels) at $E>0$, 
since there exists a closed Fermi surface at $E>0$ in the $\varepsilon^0_{\mathrm{tqD}_+}$ band,
and the semiclassical Landau quantization is expected for the closed orbit.
On the other hand, the Fermi surface in the $\varepsilon^0_{\mathrm{tqD}_-}$ band is open
and a continuous energy is expected in the $\varepsilon^0_{\mathrm{tqD}_-}$ band even 
in the presence of the magnetic field.
Quantum mechanically the Landau levels in the $\varepsilon^0_{\mathrm{tqD}_+}$ band
  couple to the continuous energy in the $\varepsilon^0_{\mathrm{tqD}_-}$ band by quantum tunneling. 
 In this subsection we show that the  coupling between the almost-localized Landau levels and
 the continuous energy cannot be neglected for the quantized energy
with the small quantum number, $n$, but it becomes small for the larger values of $n$.

In the presence of the magnetic field $\mathbf{B}$ ($\mathbf{B} = \nabla \times \mathbf{A}$,
where $\mathbf{A}$ is the vector potential), we replace $q_x$ and $q_y$ as
\begin{align}
  q_x &\rightarrow -i \hbar \frac{\partial}{\partial x} + e A_x, \\
  q_y &\rightarrow -i \hbar \frac{\partial}{\partial y} + e A_y .
\end{align}
We study the case that the uniform magnetic field $B>0$ is applied along the $z$ direction.
We take the vector potential as
\begin{equation}
 \mathbf{A}  = (-B y,0,0).
\end{equation}
Since there is 
 no explicit $x$ in Eq.~(\ref{eqeq0}), we
can write
\begin{equation}
  \Psi(x,y) = e^{i k_x x} 
  \begin{pmatrix}
  \Psi_1(y) \\ \Psi_2(y)
  \end{pmatrix} .
\end{equation}
In this case we take
\begin{equation}
   q_x \rightarrow \hbar k_x - eB y \equiv - eB \ell \bar{y},
   \label{eqqxBy}
\end{equation}
where the magnetic length $\ell$ is defined as usual,
\begin{equation}
 \ell=\sqrt{\frac{\hbar}{eB}},
\end{equation}
and  $\bar{y}$  is the dimensionless length. 
%
%
Hereafter we write $\bar{y}$ as $y$ for simplicity.

Then the equation we study is
\begin{equation}
 \begin{pmatrix}
    (\tilde{\cal{H}}_{\textrm{tqD}} )_{11}    &  (\tilde{\cal{H}}_{\textrm{tqD}} )_{12}    \\
    (\tilde{\cal{H}}_{\textrm{tqD}} )_{21}    &   (\tilde{\cal{H}}_{\textrm{tqD}} )_{22}    
  \end{pmatrix}
\begin{pmatrix}  \Psi_1({y}) \\ \Psi_2({y})  \end{pmatrix} 
 = E  \begin{pmatrix}  \Psi_1({y}) \\ \Psi_2({y})  \end{pmatrix} ,
\label{eqeq2}
\end{equation}
where
\begin{align}
   (\tilde{\cal{H}}_{\textrm{tqD}} )_{11} 
        &= w_y\sqrt{\hbar eB} \left( \frac{w_x}{w_y} {y} + \frac{\alpha_2'  \sqrt{\hbar eB}}{w_y} {y}^2  \right),   
\label{eqh11}\\
   (\tilde{\cal{H}}_{\textrm{tqD}} )_{12} 
           &= w_y\sqrt{\hbar eB} 
   \left( - \frac{d}{d {y}}  - \frac{w_x}{w_y} {y} + \frac{\alpha_2'' \sqrt{\hbar eB}}{w_y} {y}^2 \right) , 
\label{eqh12} \\
   (\tilde{\cal{H}}_{\textrm{tqD}} )_{21} 
           &= w_y\sqrt{\hbar eB} 
       \left(  \frac{d}{d {y}}  - \frac{w_x}{w_y} {y} + \frac{\alpha_2'' \sqrt{\hbar eB}}{w_y} {y}^2 \right) ,
 \label{eqh21}\\
   (\tilde{\cal{H}}_{\textrm{tqD}} )_{22} &= (\tilde{\cal{H}}_{\textrm{tqD}} )_{11} . \label{eqh22}
\end{align}%
where $w_y\sqrt{\hbar eB} = \hbar w_y/\ell$ is the energy scale for the massless Dirac fermions.
There are other dimensionless parameters,
$w_x/w_y$, $\alpha_2'\sqrt{\hbar e B}/w_y$, and $\alpha_2''\sqrt{\hbar e B}/w_y$.
We assume that $w_x/w_y$ is order of $1$ and we mainly study the case $w_x=w_y$ in this paper.
Other two dimensionless parameters are taken to be small,
i.e.,
\begin{align}
 \alpha_2'  \sqrt{\hbar e B}/w_y &\ll 1 ,\\
 \alpha_2'' \sqrt{\hbar e B}/w_y &\ll 1 .
\end{align}
We will show that the sum of these small dimensionless parameters 
($\alpha_2 \sqrt{\hbar eB}/w_y$) plays an important role in the 
quantization of energies for almost localized states in the magnetic field, but
the difference ($\tilde{\alpha}_2 \sqrt{\hbar eB}/w_y$)
is irrelevant when these parameters are small.
In other words there is another length scale $\ell \alpha_2\sqrt{\hbar eB}/w_y = \hbar \alpha_2/w_y$.


\begin{figure}[tb]
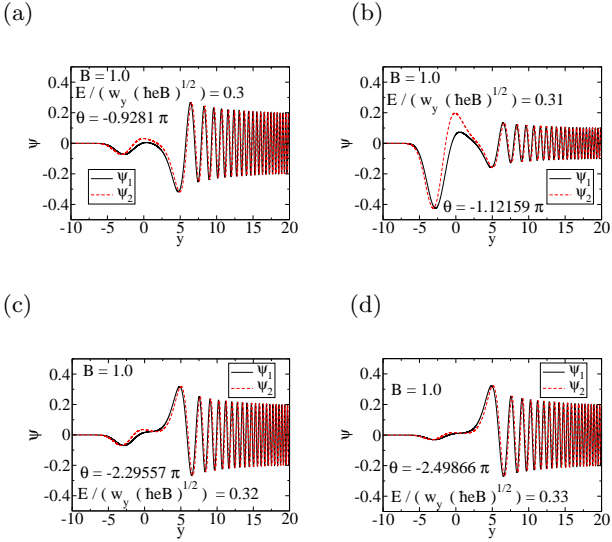

\centering
{\raggedright \ (a) \hspace{4.0cm} (b) \\ } \vspace{0.5cm} 
\includegraphics[width=0.2\textwidth]{fig2anew.eps} \hfil 
\includegraphics[width=0.2\textwidth]{fig2bnew.eps}\\ \vspace{0.5cm}
{\raggedright \ (c) \hspace{4.0cm} (d) \\ } \vspace{0.5cm}
\includegraphics[width=0.2\textwidth]{fig2cnew.eps} \hfil 
\includegraphics[width=0.2\textwidth]{fig2dnew.eps}
\caption{(Color online) 
Wave functions in the three-quarter Dirac point 
obtained numerically 
with $E/(w_y \sqrt{\hbar eB})=0.3$ (a), $0.31$ (b), $0.32$ (c), and $0.33$ (d). 
Parameters are $w_x = w_y=1$, $w_{0x}=-1$, $w_{0y}=0$,
$\alpha_2' \sqrt{\hbar eB}/w_y =\alpha_2' \sqrt{\hbar eB}/w_y=0.01$, and $B=1$.
The boundary condition at $y=20$ is taken to be Eq.~(\ref{eqbcy20}).
}
\label{fige0bx}
\end{figure}
%
\begin{figure}[tb]
\centering
{\raggedright \ (a) \hspace{4.0cm} (b) \\ } \vspace{0.5cm} 
\includegraphics[width=0.2\textwidth]{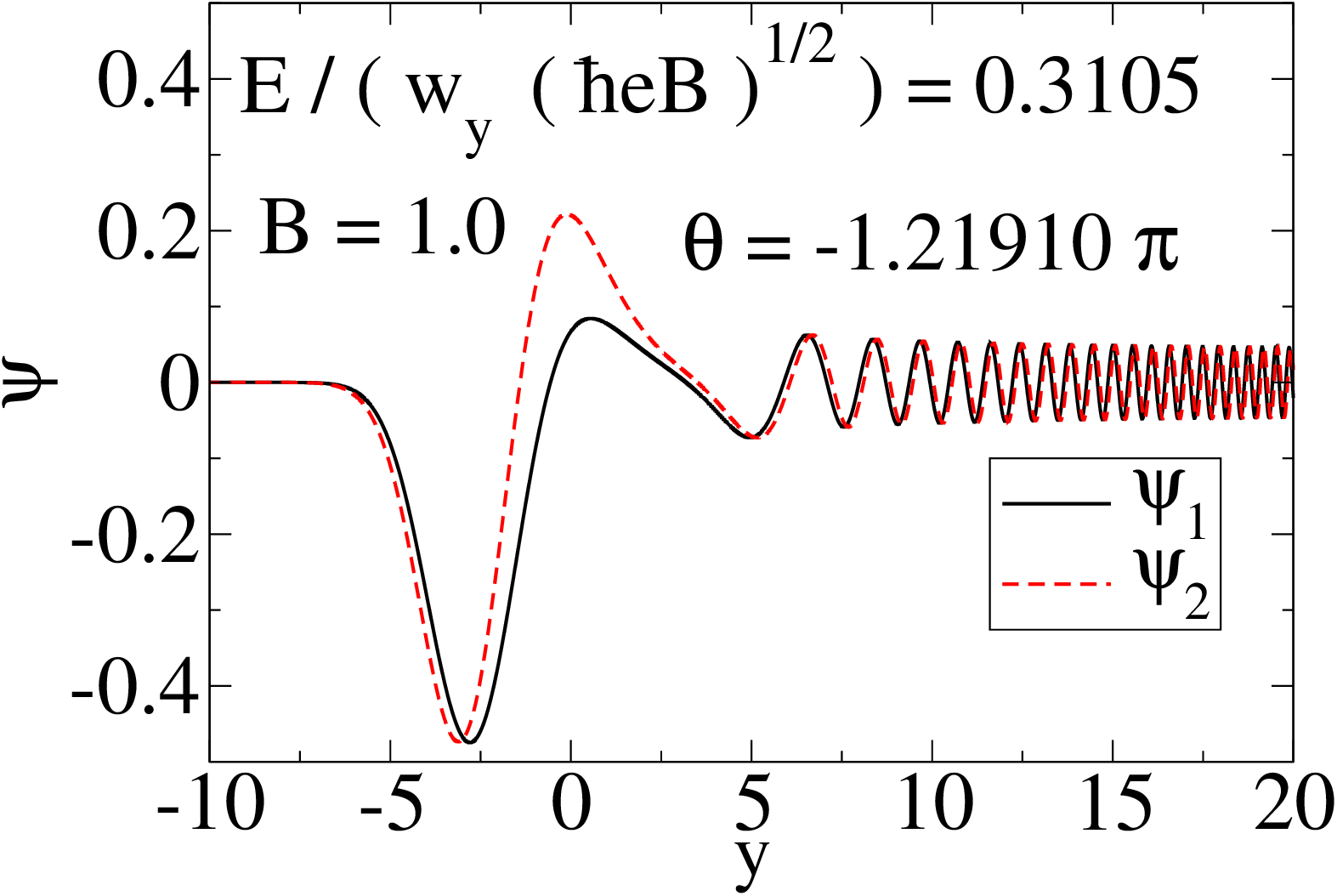}\hfil 
\includegraphics[width=0.2\textwidth]{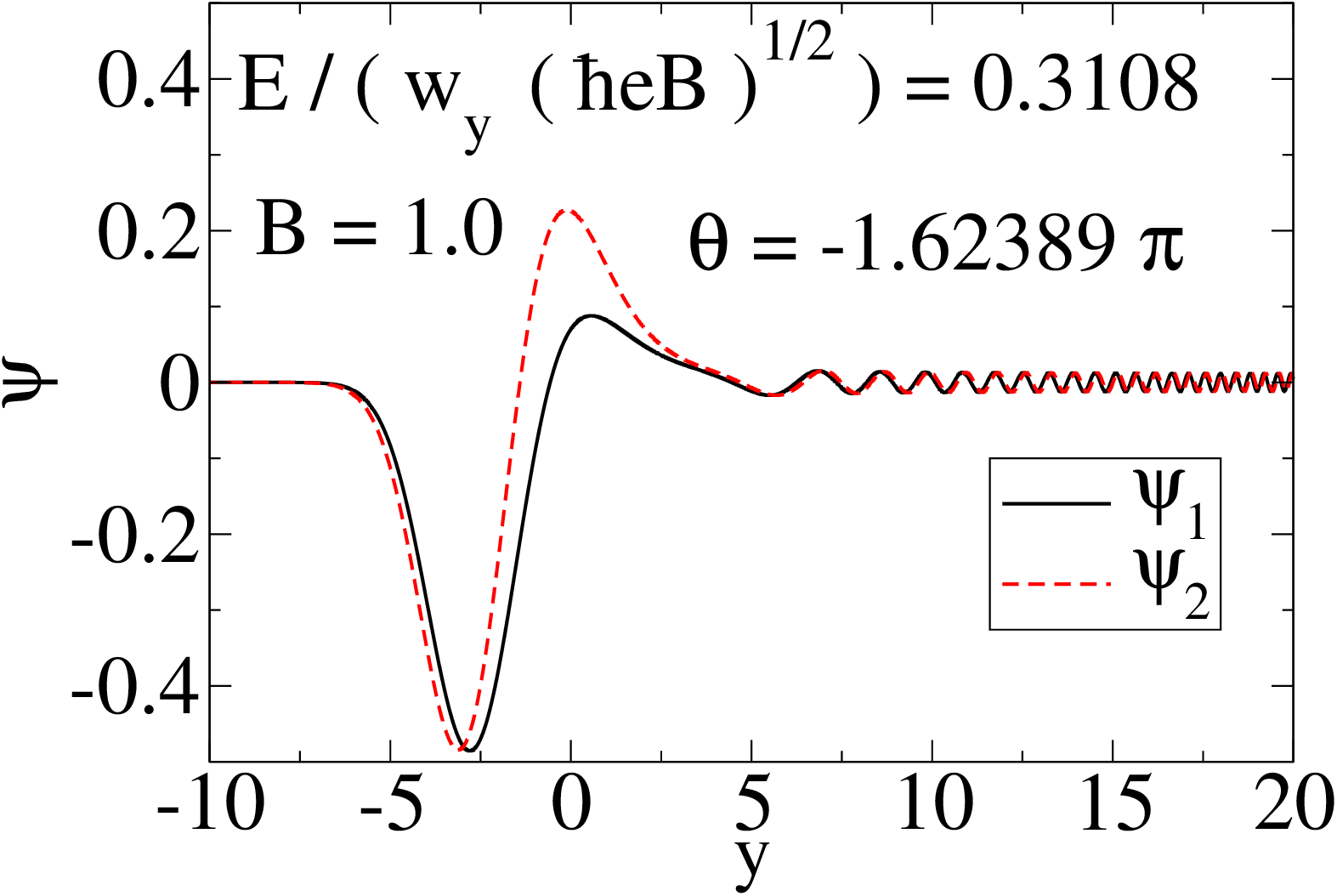}
\caption{(Color online) 
Wave functions in the three-quarter Dirac point
obtained numerically 
 with $E/(w_y \sqrt{\hbar eB})=0.3105$ (a), and $0.3108$ (b).
Parameters are $w_x = w_y=1$, $w_{0x}=-1$, $w_{0y}=0$,
$\alpha_2'\sqrt{\hbar eB}/w_y=\alpha_2''\sqrt{\hbar eB}/w_y=0.01$, and $B=1$.
The boundary condition at $y=20$ is taken to be Eq.~(\ref{eqbcy20}).
}
\label{fige0bx2}
\end{figure}
\begin{figure}[bt]
\centering
{\raggedright \ (a) \hspace{3.5cm} (b) \\ } \vspace{0.4cm}
\includegraphics[width=0.2\textwidth]{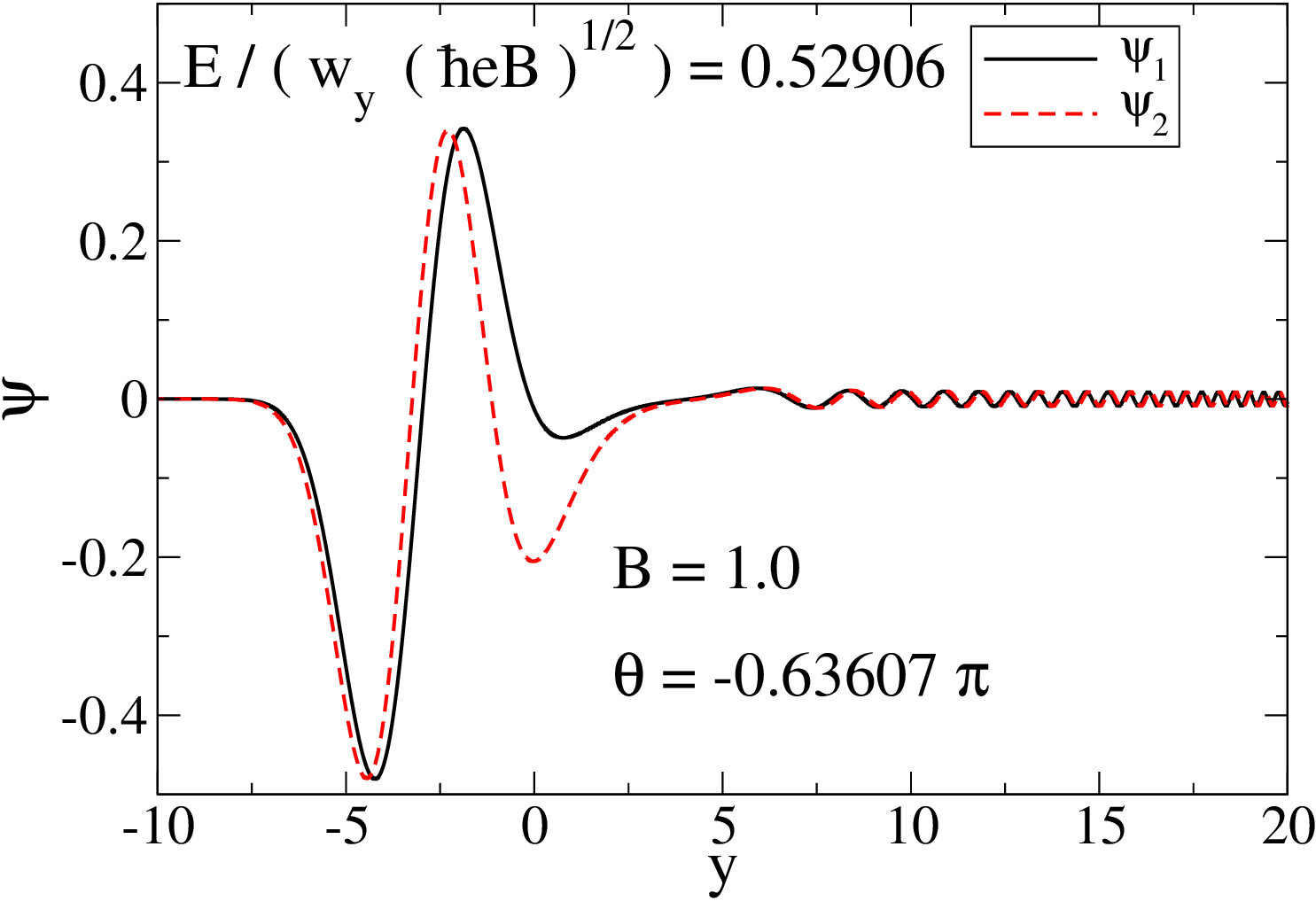} \hfil 
\includegraphics[width=0.2\textwidth]{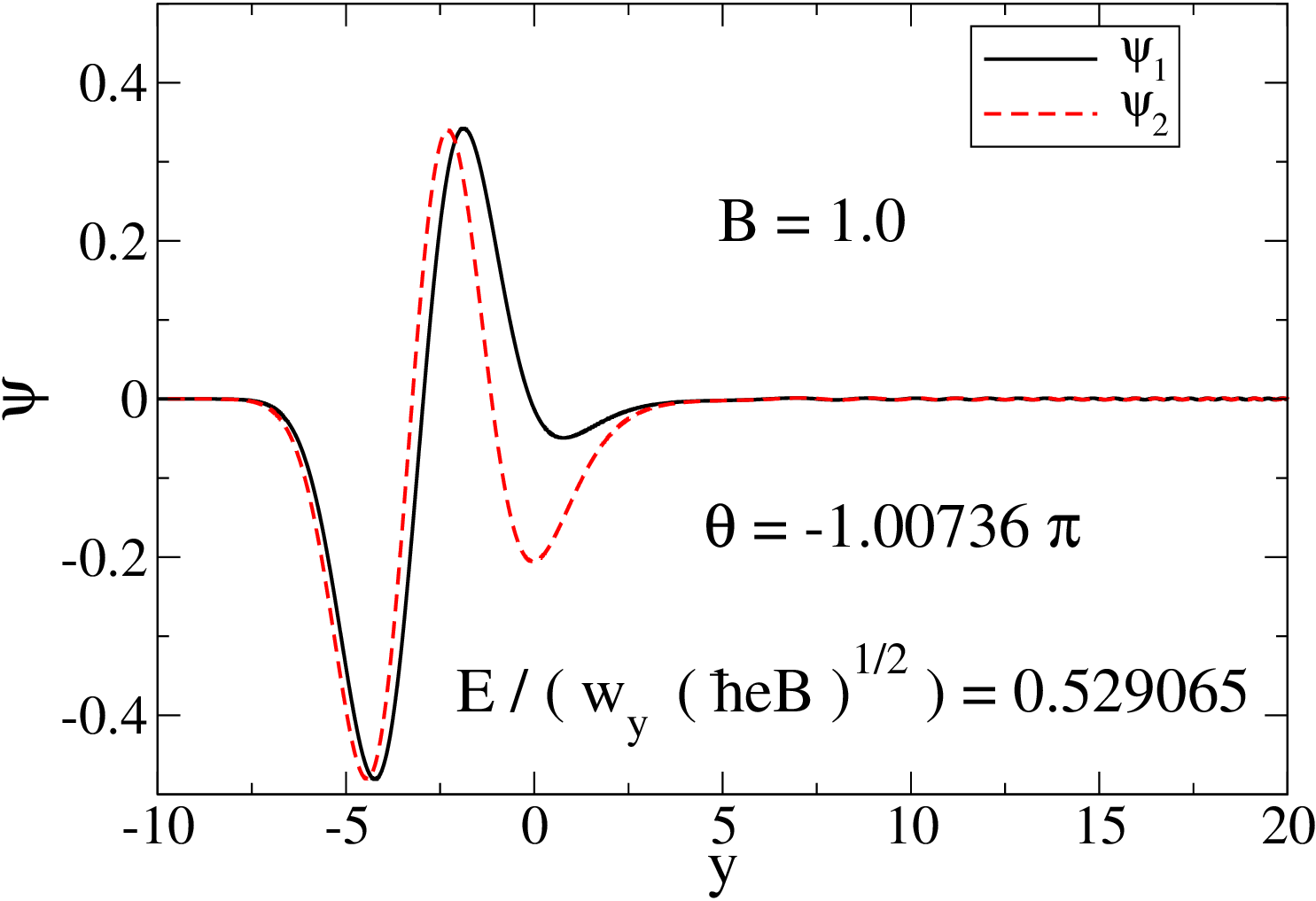}\\ \vspace{0.5cm}
{\raggedright \ (c) \\ } \vspace{0.4cm}
\hspace{0.4cm}
\includegraphics[width=0.2\textwidth]{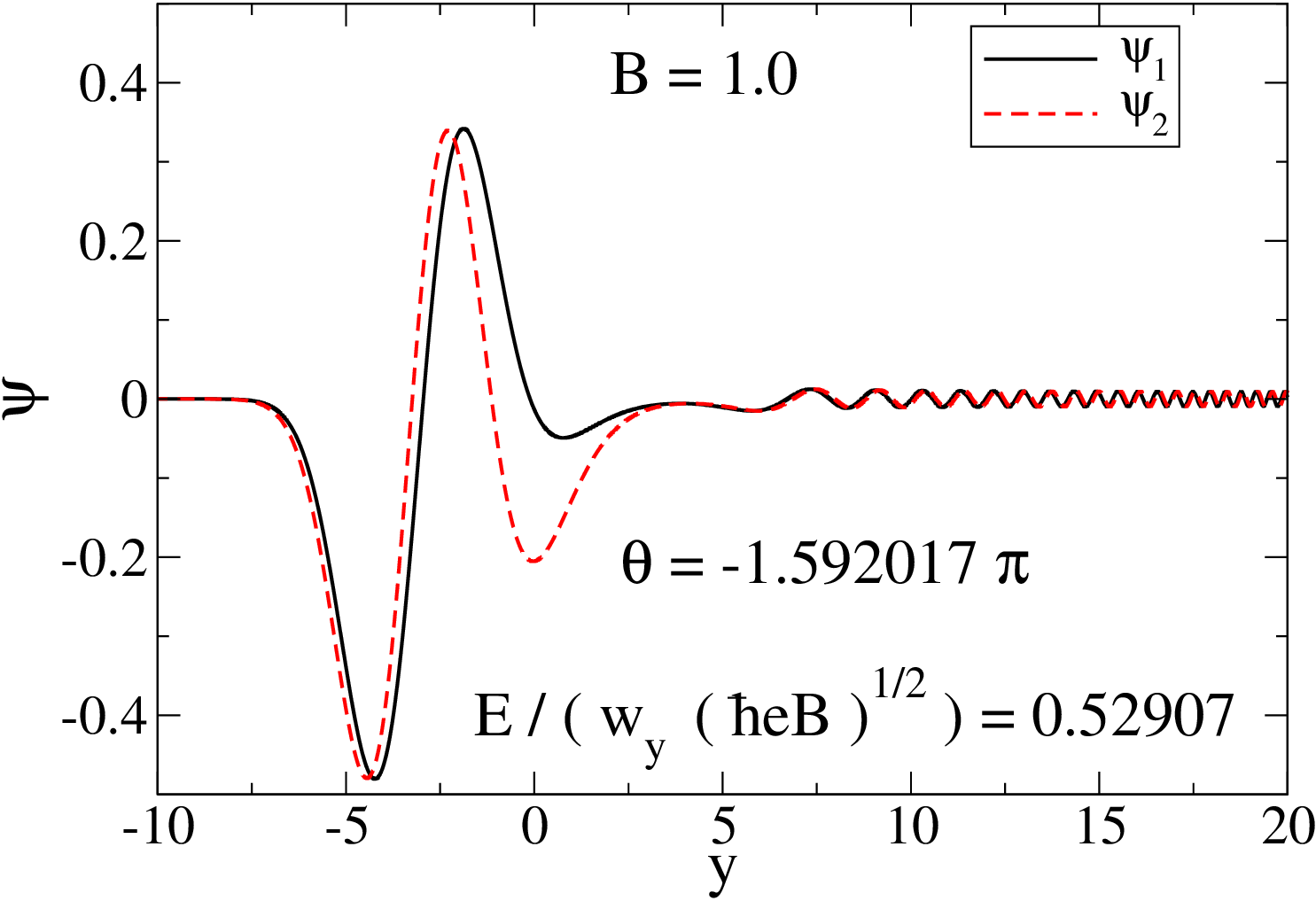}\hfill  \mbox{ } 
\caption{(Color online) 
Wave functions in the three-quarter Dirac point
obtained numerically 
with $E/(w_y \sqrt{\hbar eB})=0.52906$ (a), $0.529065$ (b),  and $0.52907$ (c).  
Parameters are $w_x = w_y=1$, $w_{0x}=-1$, $w_{0y}=0$,
$\alpha_2'\sqrt{\hbar eB}/w_y=\alpha_2''\sqrt{\hbar eB}/w_y=0.01$, and $B=1$.
The boundary condition at $y=20$ is taken to be Eq.~(\ref{eqbcy20}).
}
\label{fige1bxx}
\end{figure}
%

We seek the solution of Eq.(\ref{eqeq2}) with  $E \geq 0$
which satisfies the conditions that at $y \to -\infty$ 
\begin{align}
 \Psi_1(y) \to 0,
\label{eqymininf1}\\
 \Psi_2(y) \to 0.
 \label{eqymininf2}
\end{align}
Note that $y \to -\infty$ corresponds to $q_x \to + \infty$, as seen in Eq.~(\ref{eqqxBy}).
When $y \to +\infty$, $\Psi_1(y)$ and $\Psi_2(y)$ do not have to vanish because the lower band
becomes positive when $q_x \to -\infty$ at $B=0$ as seen in Fig.~\ref{figbandtqD}. 
Therefore, the conditions, Eqs.~(\ref{eqymininf1}) and (\ref{eqymininf2}) at $y \to - \infty$,
do not make the energy quantized. There is the solution for any value of $E$,
 but the conditions, Eqs.~(\ref{eqymininf1}) and (\ref{eqymininf2}) at $y \to - \infty$,
  make the restriction for the solutions.
 We solve the differential equations, Eq.~(\ref{eqeq2}),
 numerically by the fourth-order Runge-Kutta method in 
 this and the next subsections. We take the step size in the Runge-Kutta method to be $0.01$.
%
Since Eq.~(\ref{eqeq2}) is the real linear differential equations, the solutions can be taken as real functions,
and  the solutions multiplied by any constant values give the same solutions.
Therefore, for each value of $E$ the only adjustable parameter to obtain the solution numerically 
 by the Runge-Kutta method starting from a fixed
$y=y_{+}>0$ and decreasing $y$ is the ratio $\Psi_2(y_{+})/\Psi_1(y_{+})$. 
In this subsection we take $y_{+}=20$.
It is convenient to parametrize the ratio 
in terms of the angle $\theta$ defined by 
\begin{equation}
 \frac{\Psi_2(y=20)}{\Psi_1(y=20)} = \tan \theta .
\label{eqbcy20}
\end{equation}
The numerically obtained solution
 diverges as $y$ becomes a negative
large value, if the chosen $\theta$ is not a suitable value for the given $E$. 
Only when $\theta$ is the correct value for $E$,
the numerically obtained solution becomes zero as $y \to -\infty$.
In this way we determine $\theta$ for any given $E>0$.  
The boundary condition $\theta$ depends on the choice of $y_{+}$ 
and it does not have an important meaning. The $E$-dependence of  $\theta$, however,
is important to obtain the almost-localized state.
When  $E$ is changed continuously, 
$\theta$ changes continuously.
Note that  the energy is semi-classically quantized by the magnetic field,
since the closed Fermi surface $\varepsilon_{\textrm{tqD}_{+}}^0(\mathbf{q})$ 
exists at $B=0$. 
Quantum mechanically, 
these quantized states  in $y \lesssim 0$ couple to the continuous-energy states, which exist
mainly in $y\gtrsim 0$, 
by tunneling.  With this mixing of the states $\theta$ changes by $\pi$ in the small 
region of energy variation. 
Note that $\theta$ and $\theta + n \pi$ with integer $n$ give the same condition. 
We show some examples of the solutions 
for $0.3 \leq E /(w_y \sqrt{\hbar eB}) \leq 0.33$ in Fig.~\ref{fige0bx} and Fig.~\ref{fige0bx2}
and for $0.52906 \leq E /(w_y \sqrt{\hbar eB})  \leq 0.52907$ in Fig.~\ref{fige1bxx},
where we have normalized the wave functions numerically as
\begin{equation}
 \int_{-10}^{20} \left( |\Psi_1(y)|^2 + |\Psi_2(y)|^2 \right) dy = 1.
\end{equation}
%
%
\begin{figure}[bt]
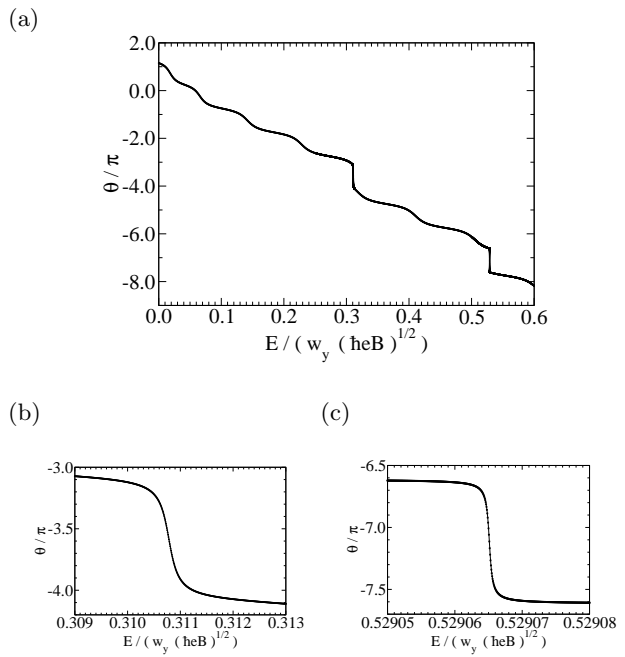

\centering
{\raggedright \ (a) \\ }
\includegraphics[width=0.33\textwidth]{fig5anew.eps}\\ \vspace{0.4cm}
{\raggedright \ (b) \hspace{3.5cm} (c) \\ } \vspace{0.4cm}
\includegraphics[width=0.2\textwidth]{fig5bnew.eps} \hfil 
\includegraphics[width=0.2\textwidth]{fig5cnew.eps}
\caption{(Color online) 
Boundary condition $\theta$ at $y=20$ (Eq.~(\ref{eqbcy20})) 
 as a function of energy for
$w_x = w_y=1$, $w_{0x}=-1$, $w_{0y}=0$,
$\alpha_2'\sqrt{\hbar eB}/w_y=\alpha_2''\sqrt{\hbar eB}/w_y=0.01$, and $B=1$.
(b) and (c) are the close-up of (a) near the energy of the almost-localized states at $y \lesssim 0$.
}
\label{figevsthetax}
\end{figure}
%

%
%
%
Nearly-localized states in  $y \lesssim 0$ exist 
at $E/(w_y \sqrt{\hbar eB}) \approx 0.3108$ and $0.529065$. 
The wave functions $(\Psi_1(y), \Psi_2(y))$ at $ E= 0.3105$ and $0.3108$ 
with the suitable boundary conditions 
have one node of $\Psi_1(y)$ and $\Psi_2(y)$ in  $y \lesssim 0$,
as seen in Fig.~\ref{fige0bx2} (a) and (b),
 and  the wave functions at $E/(w_y \sqrt{\hbar eB}) = 0.52906$, $0.529065$, and $0.52807$  
have two nodes in  $y \lesssim 0$, as seen in Fig.~\ref{fige1bxx} (a) - (c).
Therefore, $E/(w_y \sqrt{\hbar eB}) \approx 0.3108$ 
and $E/(w_y \sqrt{\hbar eB}) \approx 0.529065$  are the nearly-localized state energies with $n=1$
and $n=2$, respectively. 
Due to the tunneling  these nearly-localized states are not completely localized in the region $y \lesssim 0$,
which corresponds to the region $q_x \gtrsim 0$ in the case of $B=0$ (see Eq.~(\ref{eqqxBy})
and Fig.~\ref{figbandtqD}).
This interpretation of the nearly-localized states in three-quarter Dirac point is 
justified by plotting $\theta$ as a function of energy (Fig.~\ref{figevsthetax}).
As seen in Fig.~\ref{figevsthetax}, $\theta$ changes continuously as $E$ increases. 
When the energy is close to one of the energies of the nearly-localized states,
$\theta$ changes by $\pi$ in a narrow range of $E$.
 At $n=2$ ($E/(w_y \sqrt{\hbar eB}) \approx 0.529065$) $\theta$ 
changes in a narrower range of the energy $E$ than 
 at $n=1$ ($E/(w_y \sqrt{\hbar eB}) \approx 0.3108$). 
The narrowing of the range in $\theta$  is reasonable because the tunneling of the almost-localized state at 
$y \lesssim 0$ into the region of $y \gtrless 0$ is weaker at $n=2$ than at $n=1$. 
%
%
\begin{figure}[tb]
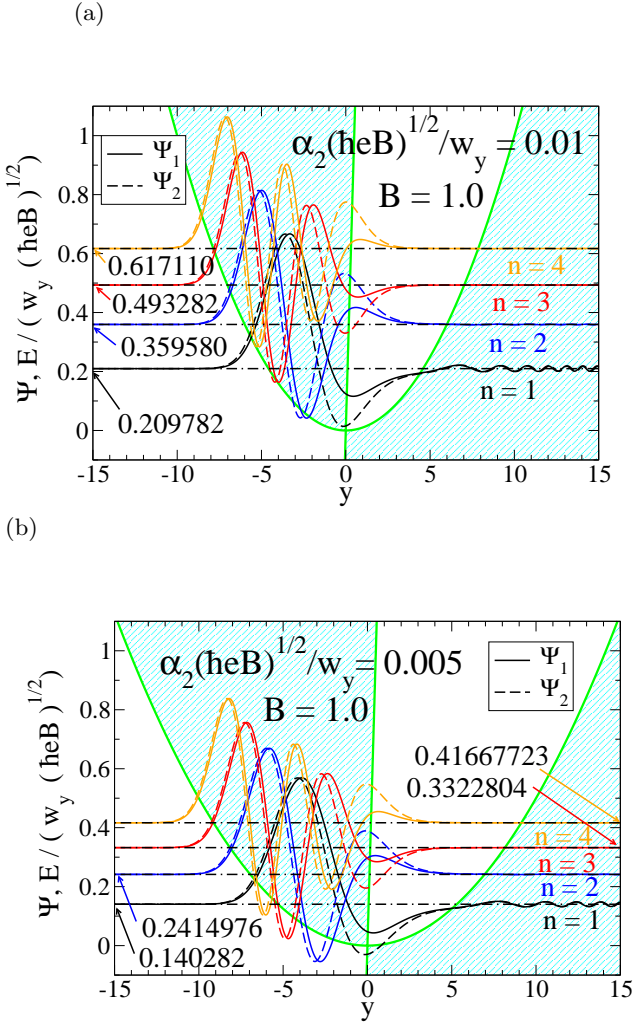

\centering
{\raggedright \hspace{0.8cm} (a)  \\ } \vspace{1.0cm} 
\includegraphics[width=0.45\textwidth]{fig6anew.eps} 
{\raggedright \hspace{0.8cm} (b)  \\ } \vspace{1.0cm} 
\includegraphics[width=0.45\textwidth]{fig6bnew.eps} 
\caption{(Color online) 
Green lines are $\varepsilon_{\mathrm{tqD}_\pm}(q_x,q_y=0)/(w_y \sqrt{\hbar eB})$ as a function of
 $y = -q_x/(\ell eB)=-q_x/\sqrt{\hbar eB}$, i.e.
$E/(w_y \sqrt{\hbar eB}) = (\alpha_2 \sqrt{\hbar eB}/w_y) y^2$ and
$E/(w_y \sqrt{\hbar eB}) = (2 w_x/w_y) y + (\tilde{\alpha}_2 \sqrt{\hbar eB}/w_y) y^2$.
Classical particles can exist in the cyan-shaded regions. 
Wave functions of almost-localized state with the quantum number $n=1$, $2$, $3$, and $4$ 
near the three-quarter Dirac point 
are plotted as  functions of $y$.  Zero of the wave functions are shifted to their energies. 
Wave functions are calculated with the boundary conditions at $y=-15$ as in section II C.
Parameters are  $B=1$, $w_x = w_y=1$, $w_{0x}=-1$, $w_{0y}=0$
in (a) and (b), and
$\alpha_2'\sqrt{\hbar eB}/w_y=\alpha_2''\sqrt{\hbar eB}/w_y=0.005$
 ($\alpha_2\sqrt{\hbar eB}/w_y=0.01$)  in (a)
 and $\alpha_2'\sqrt{\hbar eB}/w_y=\alpha_2''\sqrt{\hbar eB}/w_y=0.0025$ 
($\alpha_2\sqrt{\hbar eB}/w_y=0.005$)  in (b).
}
\label{figphiByE}
\end{figure}
%
In Fig.~\ref{figphiByE}, 
we plot $\varepsilon_{\textrm{tqD}_\pm}(q_x,q_y=0)$ at $B=0$ (Eqs.~(\ref{eq07}) and (\ref{eq09}))
with replacing $q_x \to -eB \ell y$ (Eq.~(\ref{eqqxBy})) 
divided by the energy scale of the massless Dirac fermions ($w_y\sqrt{\hbar eB}$) 
 as a function of dimensionless length $y$ for the dimensionless parameter 
$\alpha_2\sqrt{\hbar eB} /wy=0.01$ (a) and $0.005$ (b). 
In these figures we also plot  the wave functions of the almost-localized states
 at $y \lesssim 0$ with the quantum number $n=1 - 4$, which are calculated using the boundary condition
at $y=y_{-}<0$ discussed in the next subsection.
Classically, electrons can exist in the cyan-shaded regions in Fig.~\ref{figphiByE}, and they can
exist only by the quantum tunneling effect in the white regions.
For the larger energy (larger quantum number $n$) the width and the hight of the classically-forbidden
region (white region in Fig.~\ref{figphiByE}) is larger. As a result the tunneling of the almost localized state
with the larger quantum number at $y \lesssim 0$ into the $y \gtrsim 0$ 
region becomes smaller. 
%
Therefore, the numerical solutions of the bound states $n \geq 3$ are difficult to obtain by using  
the boundary condition at $y=y_{+} >0$,
 Eq.~(\ref{eqbcy20}), since $\theta$ changes by $\pi$ in a very narrow region
in energy.
On the other hand, the almost-localized state with the quantum number $n=1$ couples strongly  to
the continuous states at $y \gtrsim 0$  as seen in Fig.~\ref{figevsthetax}(b),
and the energy of 
the almost-localized state is ``broadened''.

In the next subsection we use the boundary condition 
at $y<0$ to obtain the energy of the bound states. 
%
\subsection{numerical results of  energy at $B > 0$, using boundary condition at $y<0$}
\label{sectionbc2}
%
%
\begin{figure}[tb]
\begin{center}
\includegraphics[width=0.45\textwidth]{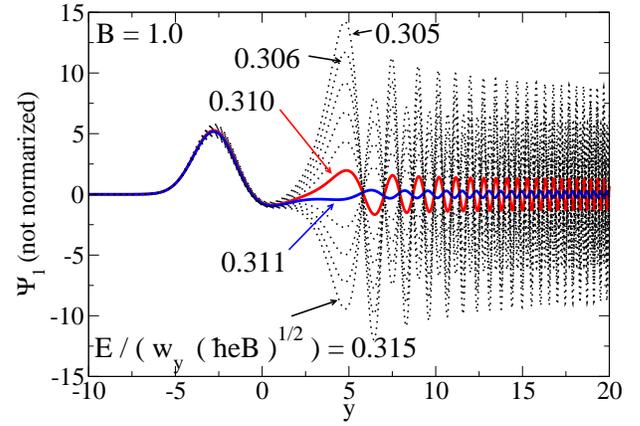}
\end{center}
\caption{(Color online) 
Wave function $\Psi_1(y)$ in the three-quarter Dirac point
obtained numerically by using the boundary condition at $y=-10$. 
Parameters are $w_x = w_y=1$, $w_{0x}=-1$, $w_{0y}=0$,
$\alpha_2'\sqrt{\hbar eB}/w_y=\alpha_2''\sqrt{\hbar eB}/w_y=0.01$, 
and $B=1$. We take several values of $E$, and we find that an eigenvalue
for the nearly localized state at $y<0$
exists in the region $0.310 < E/(w_y \sqrt{\hbar eB}) < 0.311$.
}
\label{figen1x}
\end{figure}
%
As shown in the previous subsection, it is difficult to obtain the energy
of the almost-localized states at $y \lesssim 0$ with a large quantum number $n$
 in  Eq.(\ref{eqeq2}) by using the boundary condition at $y = y_{+} >0$,
since the  boundary condition changes in a very narrow region and 
 the energy
of the almost-localized states at $y \lesssim 0$ may be overlooked.
Therefore, we try to obtain the energy by using the boundary conditions at $y<0$.
We study the solutions of Eq.(\ref{eqeq2}) at $y \to -\infty$, assuming
\begin{equation}
  \Psi_j(y) = c_j(y) e^{-g(y)},
\end{equation}
($j=1,2$) and 
\begin{align}
  \frac{d \Psi_j}{dy} &= \left(- \frac{dg(y)}{dy} c_j(y) +\frac{d c_j(y)}{d y} \right) e^{-g(y)} \nonumber \\
                      &\sim - \frac{dg(y)}{dy} c_j  e^{-g(y)} ,
\end{align}
as $y \to -\infty$.
Then we obtain the equation
\begin{equation}
\begin{pmatrix}
    (\tilde{\cal{H}}_{\textrm{tqD}} )_{11}    &  (\tilde{\cal{F}}_{\textrm{tqD}} )_{12}    \\
    (\tilde{\cal{F}}_{\textrm{tqD}} )_{21}   &   (\tilde{\cal{H}}_{\textrm{tqD}} )_{22}    
  \end{pmatrix}
\begin{pmatrix}  \Psi_1(y) \\ \Psi_2(y)  \end{pmatrix} 
 \approx E  \begin{pmatrix}  \Psi_1(y) \\ \Psi_2(y)  \end{pmatrix} ,
\label{eqapprox}
\end{equation}
where $(\tilde{\cal{H}}_{\textrm{tqD}} )_{11}$ and $(\tilde{\cal{H}}_{\textrm{tqD}} )_{22}$ 
are given in Eqs.~(\ref{eqh11}) and (\ref{eqh22}) and
\begin{align}
   (\tilde{\cal{F}}_{\textrm{tqD}} )_{12} &=
 w_y \sqrt{\hbar eB} \left( \frac{d g}{dy} -\frac{w_x}{w_y} y + \frac{\alpha'' \sqrt{\hbar eB}}{w_y} y^2 \right) ,
\label{eqh12b} \\
   (\tilde{\cal{F}}_{\textrm{tqD}} )_{21} &= 
 w_y \sqrt{\hbar eB} \left( - \frac{d g}{dy} -\frac{w_x}{w_y} y + \frac{\alpha'' \sqrt{\hbar eB}}{w_y} y^2 \right) .
\label{eqh21b}
\end{align}
The nontrivial solution exists when the condition
\begin{equation}
 \det
\begin{pmatrix}
    (\tilde{\cal{H}}_{\textrm{tqD}} )_{11}  -E  &  (\tilde{\cal{F}}_{\textrm{tqD}} )_{12}    \\
    (\tilde{\cal{F}}_{\textrm{tqD}} )_{21}   &   (\tilde{\cal{H}}_{\textrm{tqD}} )_{22}    -E
  \end{pmatrix}
=0,
\end{equation}
is fulfilled, i.e.,
\begin{align}
  \left(  \frac{d g}{dy} \right)^2 &=
  \left( \frac{\alpha_2  \sqrt{\hbar eB}}{w_y} y^2 - \frac{E}{w_y \sqrt{\hbar eB}} \right) \nonumber \\
& \times  \left(- \frac{2 w_x}{w_y} y - \frac{\tilde{\alpha}_2 \sqrt{\hbar eB}}{w_y} y^2 
 + \frac{E}{w_y\sqrt{\hbar eB}} \right),
\end{align}
In the simple case that $\alpha_2 >0$, $\tilde{\alpha}_2=0$ (i.e., $\alpha'=\alpha''=\alpha_2/2$), and large $|y|$,
we can neglect terms proportional to $E$.
Then the approximate solution is
\begin{equation}
 g(y) \sim \pm \frac{2 \sqrt{2 \alpha_2 w_x \sqrt{\hbar eB}}}{5 w_y}  (-y)^{\frac{5}{2}} + \textrm{const.}
\end{equation}
The solution which does not diverge at $y \to -\infty$ is obtained as
\begin{equation}
 \begin{pmatrix} \Psi_1 \\ \Psi_2 \end{pmatrix}
  \sim \exp \left( - \frac{2 \sqrt{2 \alpha_2 w_x \sqrt{\hbar eB}}}{5 w_y}  (-y)^{\frac{5}{2}} \right) 
   \begin{pmatrix} c_1 \\ c_2 \end{pmatrix} .
\label{eqf12}
\end{equation}

Inserting Eq.~(\ref{eqf12}) into Eq.~(\ref{eqapprox}) we obtain
the approximate boundary condition at $y \to -\infty$ as
\begin{align}
 \frac{\Psi_1(y)}{\Psi_2(y)} 
&\sim \frac{c_1(y)}{c_2(y)} \nonumber \\ 
&\sim \frac{w_x+{\alpha_2'} B\sqrt{\hbar eB} y}{w_x-{\alpha_2''} \sqrt{\hbar eB} y 
 -\sqrt{2\alpha_2 w_x \sqrt{\hbar eB (-y)}}} .
\end{align}
With this boundary conditions at $y=y_{-}=-10.0$  we solve the differential equation
Eq.~(\ref{eqeq2}) numerically in the Runge-Kutta method with increasing $y$.
When we take $E$ to be one of the correct values of the Landau levels,
the wave function is nearly localized at $y \lesssim 0$ and tunnels to $y>0$ very little. On the other hand, 
if we take the different value of $E$, 
the wave function becomes large as $y$ is increased at $y>0$, 
although it does not diverge. 
As shown in Fig.~\ref{figen1x}, the wave function in the region $y>0$ 
calculated numerically 
with the boundary condition at $y_{-}=-10$ becomes small only when we take the correct eigenvalue
$0.310 < E < 0.311$.  This value is consistent with the $n=1$ eigenvalue obtained
numerically with the boundary condition at $y_{+}=20$ (Fig.~\ref{fige0bx2}).
 We also check numerically that the solution is not sensitive to the boundary condition; 
 numerically the same result is obtained even when we take $\Psi_1=0$ and $\Psi_2 \neq 0$ at $y=y_{-}=-10$.
 The independence on the boundary condition can be understood as follows. 
 As seen in section II B, 
the coupling between the nearly-localized state at $y \lesssim 0$ and the continuous state as $y \gtrsim 0$ 
is small for $n \geq 2$. In section II~B we first fixed the energy and obtain the wave functions
not divergent at $y \to -\infty$ 
by changing the boundary condition at $y_{+}>0$ 
($\theta$ at $y_{+}=20$). 
In this section we first take the approximate boundary condition at $y_{-}=-10$, and 
obtain the energy which gives the smallest amplitude of oscillations of the wave function at $y>0$.
Even though the boundary condition is not exact, suitable linear combination of the  
 nearly-localized state at $y \lesssim 0$ and continuous state as $y \gtrsim 0$ may give the 
non-divergent solution  with the given boundary condition at
$y=y_{-}$, if the energy is the correct energy of the  nearly-localized state at $y \lesssim 0$.

\begin{figure}[tb]
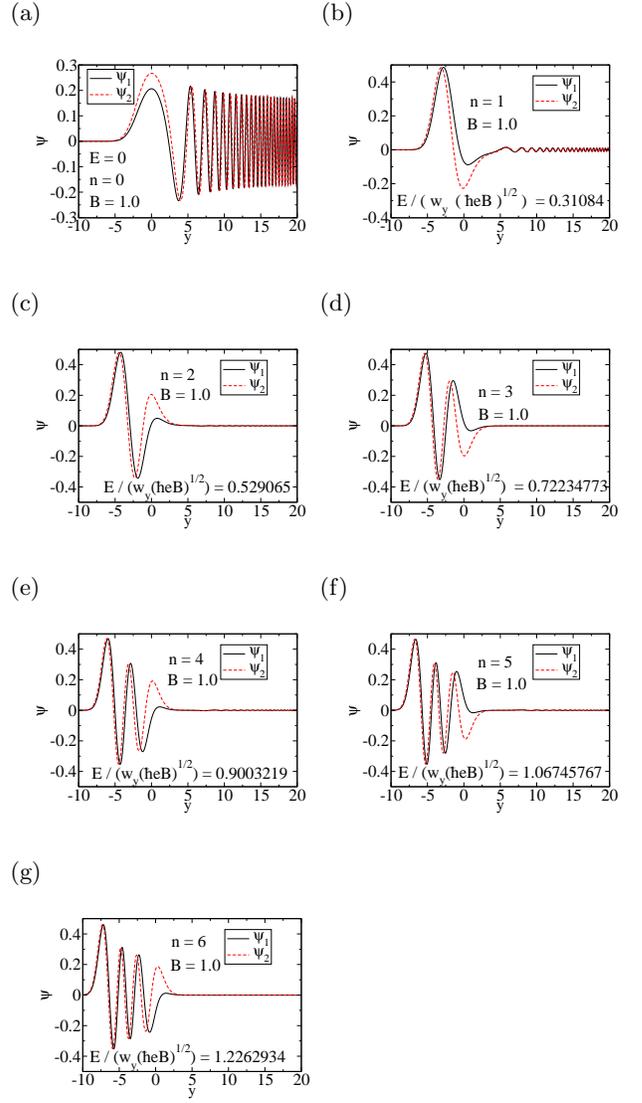

\centering
{\raggedright \ (a) \hspace{3.5cm} (b) \\ } \vspace{0.4cm}
\includegraphics[width=0.2\textwidth]{fig8anew.eps} \hfil 
\includegraphics[width=0.2\textwidth]{fig8bnew.eps}\\ \vspace{0.5cm}
{\raggedright \ (c) \hspace{3.5cm} (d) \\ } \vspace{0.4cm}
\includegraphics[width=0.2\textwidth]{fig8cnew.eps} \hfil 
\includegraphics[width=0.2\textwidth]{fig8dnew.eps}\\ \vspace{0.5cm}
{\raggedright \ (e) \hspace{3.5cm} (f) \\ } \vspace{0.4cm}
\includegraphics[width=0.2\textwidth]{fig8enew.eps} \hfil 
\includegraphics[width=0.2\textwidth]{fig8fnew.eps}\\ \vspace{0.5cm}
{\raggedright \ (g)  \\ } \vspace{0.4cm}
\hspace{0.4cm} \includegraphics[width=0.2\textwidth]{fig8gnew.eps} \hfill  \mbox{ }
\caption{(Color online) 
Wave functions of 
nearly localized eigenstates at $y<0$ 
with quantum number (a) $n=0$,  (b) $n=1$, (c) $n=2$,  (d) $n=3$, (e) $n=4$,  (f) $n=5$, and
(g) $n=6$  
 in the three-quarter Dirac point obtained numerically with the boundary condition at $y=-10$. 
Parameters are $w_x = w_y=1$, $w_{0x}=-1$, $w_{0y}=0$,
$\alpha_2'\sqrt{\hbar eB}/w_y=\alpha_2''\sqrt{\hbar eB}/w_y=0.01$, and $B=1$.
}
\label{fige0}
\end{figure}

In Fig.~\ref{fige0} we show the wave functions for nearly-localized states with 
quantum numbers $n=0$ -- $6$.
%
For $n=0$, i.e. 
 $E=0$, both components of the wave function
have a broad peak at $y=0$,
although each component of the wave functions is not small at $y>0$, as shown in Fig.~\ref{fige0}(a).
The oscillation of the wave function at $y>0$ can be understood as the continuous energy states at $y>0$.
Since the upper band touches  the lower band at the three-quarter Dirac point
without the boundary barrier, the nearly-localized state at $y<0$ goes through to the region $y>0$.
We will discuss the $n=0$ state in the next section.

The eigenstate for $n \geq 1$ is obtained by taking the suitable value of $E$, which minimize the amplitude
of oscillation of the wave function in the region $y>0$.
We find the tunneling through the barrier is smaller as $n$ becomes larger, 
as we have discussed in the previous subsection.

\begin{figure}[tb]
\centering
\includegraphics[width=0.4\textwidth]{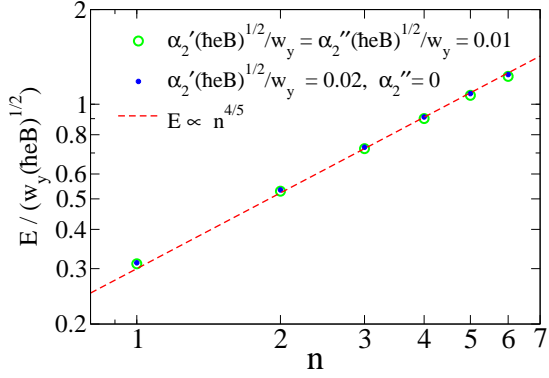}
\caption{(Color online) 
Energy as a function of the quantum number $n$ for the three-quarter Dirac point. 
Parameters are $w_x = w_y=1$, $w_{0x}=-1$, $w_{0y}=0$,
 and $B=1$. We take two choices of parameters giving the same 
$\alpha_2$, $\alpha_2'\sqrt{\hbar eB}/w_y=\alpha_2''\sqrt{\hbar eB}/w_y=0.01$ and
$\alpha_2'\sqrt{\hbar eB}/w_y=0.02$, $\alpha_2''=0$. 
The obtained values of the energy is well fitted by the red broken line ($E \propto 0.3 n^{\frac{4}{5}}$). 
}
\label{figEvsn}
\end{figure}

\begin{figure}[tb]
\hspace{1cm}
\begin{center}
\includegraphics[width=0.4\textwidth]{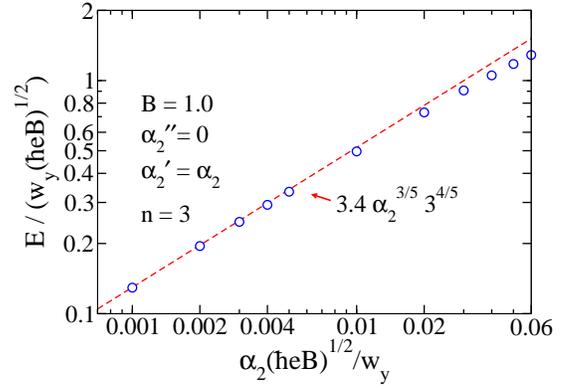}
\end{center}
\caption{(Color online) 
Energy with the quantum number $n=3$ as a function of $\alpha_2$ for the three-quarter Dirac point. 
Parameters are $w_x = w_y=1$, $w_{0x}=-1$, $w_{0y}=0$,
$\alpha_2''=0$, and $B=1$.
}
\label{figEvsalpha}
\end{figure}

\begin{figure}[tb]
\begin{center}
\vspace{5mm}
\includegraphics[width=0.4\textwidth]{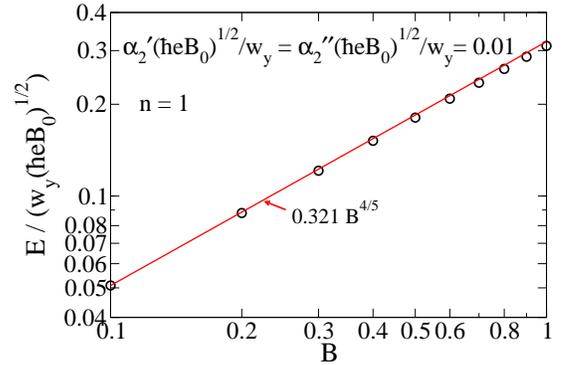}
\end{center}
\caption{(Color online) 
Dimensionless energy ($E/(w_y\sqrt{\hbar e B_0})$ with $B_0=1$) 
as a function of magnetic field $B$ at the three quarter Dirac point. 
It is well fitted as a function of $B^{\frac{4}{5}}$.
}
\label{figEvsB}
\end{figure}

We also calculate the energy as a function of quantum number $n$ with different choice of parameters
$\alpha_2'\sqrt{\hbar eB}/w_y=0.02$ and $\alpha_2''=0$ from these used 
in Fig.~\ref{fige0} ($\alpha_2'\sqrt{\hbar eB}/w_y=\alpha_2''\sqrt{\hbar eB}/w_y=0.01$). 
We plot the energy as a function of $n$  in Fig.~\ref{figEvsn}. We obtain 
 \begin{equation}
  E_n \propto n^{\frac{4}{5}}.
\label{eqpropto} 
\end{equation}
In Figs.~\ref{figEvsalpha} and \ref{figEvsB} 
we plot the energy as a function of $\alpha_2$ and $B$, respectively. We obtain 
\begin{equation}
 E_n \propto \alpha_2^{\frac{3}{5}} (n B)^{\frac{4}{5}}.
 \label{eqquantizedE}
\end{equation}
We have previously obtained $n$ and $B$ dependence 
at the three-quarter Dirac point (Eq.~(\ref{eqquantizedE}))
 in the tight-binding model of $\alpha$-(BEDT-TTF)$_2$I$_3$
at the critical pressure\cite{Kishigi2017}. 
%
%
%
%
%
%
%
\subsection{analytical study with approximation in the
magnetic-field- and $\alpha_2$-dependence of the Landau levels at the three-quarter Dirac point}
In this subsection we give the analytical derivation of Eq.~(\ref{eqquantizedE}).
Taking a sum and a difference, we obtain  the equations
\begin{align}
&
 \left[ \frac{\alpha_2\sqrt{\hbar eB}}{w_y} {y}^2 -\frac{E}{w_y\sqrt{\hbar eB}} \right] 
\left(\Psi_1+\Psi_2\right) \nonumber \\
&+ \frac{d}{d{y}}  \left( \Psi_1 - \Psi_2 \right) = 0  ,
\label{eqeqanalA} \\
&\left[ \frac{ 2 w_x }{w_y} y 
+  \frac{\tilde{\alpha}_2\sqrt{\hbar eB}}{w_y} {y}^2 - \frac{E}{w_y \sqrt{\hbar eB}} \right]
\left( \Psi_1-\Psi_2 \right) \nonumber \\
&-  \frac{d}{d{y}} \left( \Psi_1 + \Psi_2 \right) = 0 .
\label{eqeqanalB}
\end{align}
In the three-quarter Dirac case studied in this paper 
the term proportional to ${y}$  in Eq.~(\ref{eqeqanalA})  does not exist and
 the term proportional to ${y}^2$ in Eq.~(\ref{eqeqanalA}) cannot be neglected, while 
the term proportional to ${y}^2$ in Eq.~(\ref{eqeqanalB}) can be neglected.
Then there appear  dimensionless parameters 
 $\alpha_2 \sqrt{\hbar eB}/w_y$ and $w_x/w_y$.
The energy depends not only the energy scale
$w_y\sqrt {\hbar eB}$ but also these dimensionless parameters. Therefore, we may expect 
\begin{equation}
E \propto \left( w_y \sqrt{\hbar e B} \right) \left(\frac{\alpha_2 \sqrt{\hbar eB}}{w_y}\right)^{\beta}
\left( \frac{w_x}{w_y} \right) ^{\eta} n^\delta,
\label{eqdimE}
\end{equation}
where $n$ is the quantum number of the almost localized state at $y \lesssim 0$.
We determine the exponents, $\beta$, $\eta$ and $\delta$. We take
\begin{equation}
  \beta > 0 ,
\end{equation}
in order to obtain $E \to 0$ as $\alpha_2 \to 0$.
The almost-localized state has
the finite absolute value of $|\Psi_1+\Psi_2|$ 
 in the region
\begin{equation}
-{y}_0 \lesssim {y} \lesssim 0,
\end{equation}
 and it is  exponentially small in the region
\begin{equation}
 {y} \lesssim -{y}_0,
\end{equation}
where the dimensionless length ${y}_0$ is determined by the equation
\begin{equation}
 \frac{\alpha_2 \sqrt{\hbar eB}}{w_y} y_0^2 = \frac{E}{w_y \sqrt{\hbar e B}} \sim  
 \left(\frac{\alpha_2 \sqrt{\hbar eB}}{w_y}\right)^{\beta}
\left( \frac{w_x}{w_y} \right) ^{\eta} n^{\delta}.
\end{equation}  
Then $y_0$ depends on the dimensionless parameters as
\begin{equation}
 y_0  \sim \left(\frac{\alpha_2 \sqrt{\hbar eB}}{w_y}\right)^{\frac{\beta-1}{2}}
\left(\frac{w_x}{w_y} \right)^{\frac{\eta}{2}} n^{\frac{1}{2} \delta}.
\end{equation}
We expect
\begin{equation}
\left\langle \left| \frac{d}{dy} (\Psi_1-\Psi_2) \right| \right\rangle 
\sim
 \frac{2 c n}{ {y}_0} 
\Bigl\langle \sqrt{|\Psi_1-\Psi_2|^2} \Bigr\rangle ,
\label{eqapprox0}
\end{equation} 
where $\langle \cdots \rangle$ is the spacial average in ${y}_0 \lesssim {y} \lesssim 0$
and $c$ is a dimensionless constant of order $1$.
This approximation is not justified for small $n$.
However, we may consider that   $(\Psi_1-\Psi_2)$ changes sign $n$ times in the length of $y_0$,
i.e.,
 $(\Psi_1-\Psi_2)$ changes from $\pm c
\left\langle \sqrt{|\Psi_1-\Psi_2|^2} \right\rangle$ to
 $\mp c 
\left\langle \sqrt{|\Psi_1-\Psi_2|^2} \right\rangle$ periodically in the half period ($y_0/n$).
 Approximating the oscillation of $(\Psi_1-\Psi_2)$ by a triangle wave, we obtain 
 Eq.~(\ref{eqapprox0}).
This crude approximation will give an approximate dependence  on $n$ and $y_0$ 
in Eq.~(\ref{eqapprox0}) in the limit of  $n \gg 1$.
With this approximation  we obtain 
\begin{equation}
 \frac{\Bigl\langle \sqrt{|\Psi_1-\Psi_2|^2} \Bigr\rangle}{\Bigl\langle \sqrt{|\Psi_1+\Psi_2|^2} \Bigr\rangle}
 \sim \left(\frac{\alpha_2\sqrt{\hbar eB}}{w_y} \right)^{\frac{3\beta-1}{2}} 
\left( \frac{w_x}{w_y} \right)^{\frac{3}{2}\eta} (2 c n)^{\frac{3\delta-2}{2}},
\label{eqbeta1}
\end{equation} 
by taking the spacial average in Eq.~(\ref{eqeqanalA}).
Next, we examine Eq.~(\ref{eqeqanalB}) in the same way. The second term and
the third term in the coefficient of $\Psi_1-\Psi_2$
 in Eq.~(\ref{eqeqanalB}), which depend on the dimensionless parameter as
$\left(\alpha_2 \sqrt{\hbar eB}/w_y\right)^{\beta}$, can be neglected with respect to the first
term  in the coefficient of $\Psi_1-\Psi_2$,
since we study the case
\begin{equation}
 \frac{\alpha_2 \sqrt{\hbar eB}}{w_y} \ll \frac{w_x}{w_y}.
\end{equation}
Then we obtain
\begin{equation}
 \frac{\Bigl\langle \sqrt{|\Psi_1-\Psi_2|^2} \Bigr\rangle}{\Bigl\langle \sqrt{|\Psi_1+\Psi_2|^2} \Bigr\rangle}
 \sim \left(\frac{\alpha_2\sqrt{\hbar eB}}{w_y} \right)^{1-\beta}  \left( \frac{w_x}{w_y} \right)^{1-\eta} 
(2 c n)^{-\delta-1}.
 \label{eqbeta2}
\end{equation}
Comparing Eq.~(\ref{eqbeta1}) and Eq.~(\ref{eqbeta2}), we obtain 
\begin{equation}
 \beta=\frac{3}{5},
\end{equation}
\begin{equation}
 \eta=-\frac{2}{5},
\end{equation}
and
\begin{equation}
 \delta=\frac{4}{5}.
\end{equation}
Inserting these exponents in Eq.~(\ref{eqdimE}), we obtain
\begin{equation}
 E \sim w_x^{-\frac{2}{5}}w_y^{\frac{4}{5}} \alpha_2^{\frac{3}{5}} (n \hbar eB)^{\frac{4}{5}}.
 \label{eqeq22x}
\end{equation} 
In Appendix we give a simpler derivation of  Eq.~(\ref{eqeq22x}).

\begin{figure}[bt]
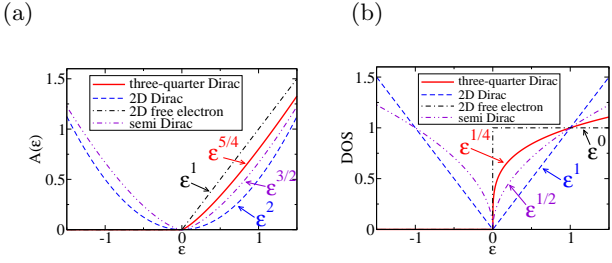

\centering
{\raggedright \ (a) \hspace{4.0cm} (b) \\ } \vspace{0.5cm} 
\includegraphics[width=0.2\textwidth]{fig12a.eps} \hfil 
\includegraphics[width=0.2\textwidth]{fig12b.eps}
\caption{(Color online) 
Schematic plot of the areas of the Fermi surface and the density of states as a function of energy
for the three-quarter Dirac fermion, massless 2D Dirac fermion, 2D free fermion, and
2D semi Dirac fermion\cite{Hasegawa2006,Dietl2008}. The density of states are scaled to be 1 at $\epsilon = 1$. 
}
\label{figtilted2}
\end{figure}

This result is consistent with the result obtained by the semiclassical quantization rule
in the previous paper\cite{Kishigi2017},
in which the energy is quantized as
\begin{equation}
A(E_n) \propto (n+ \gamma) B,
\end{equation}
where  $\gamma$ is a phase factor 
($\gamma=1/2$ for 2D free electrons and semi-Dirac fermions
and $\gamma=0$ for Dirac fermions and three-quarter Dirac fermions) and 
$A(\varepsilon)$ is the area of the Fermi surface in the 2D $\mathbf{k}$-space 
at $B=0$ with the Fermi energy $\varepsilon$. 
The area, $A(\epsilon)$, and the density of states, $D(\epsilon)$, are related by 
\begin{equation}
 \frac{1}{4\pi^2} \frac{d A(\varepsilon)}{d \varepsilon} = D(\varepsilon). 
\end{equation}
We plot $A(\varepsilon)$ and $D(\varepsilon)$ in Fig.~\ref{figtilted2}.
In the three-quarter Dirac case,  we have obtained\cite{Kishigi2017}
\begin{equation}
 A(\varepsilon) \propto \alpha_2^{-\frac{3}{4}} \varepsilon^{\frac{5}{4}},
\end{equation}
in the limit $\epsilon \to 0$,
and 
\begin{equation}
 E_n \propto \alpha_2^{\frac{3}{5}} (nB)^{\frac{4}{5}}.
\end{equation}

%
\section{finite energy gap and $n=0$ state}
In this section we study the $n=0$ state by introducing the energy gap 
in the three-quarter Dirac point, which may be caused by a difference of the site energy
in two sublattices,
\begin{equation}
 \mathcal{H}_{\textrm{tqD}}^{\prime} =
 \mathcal{H}_{\textrm{tqD}} + 
\begin{pmatrix}
 \Delta & 0 \\
0         & -\Delta
\end{pmatrix},
\label{eqtqDgap}
\end{equation}
where $2 |\Delta|$ is the energy gap at the three-quarter Dirac point. Note that
the minimum of the upper band is not at the three-quarter Dirac point ($\mathbf{q}=0$)
and the minimum energy of the upper band is not $|\Delta|$.
Then the equation we study at $B \neq 0$ is
\begin{align} 
\begin{pmatrix}
    (\tilde{\cal{H}}_{\textrm{tqD}} )_{11} + \Delta   &  (\tilde{\cal{H}}_{\textrm{tqD}} )_{12}    \\
    (\tilde{\cal{H}}_{\textrm{tqD}} )_{21}    &   (\tilde{\cal{H}}_{\textrm{tqD}} )_{22} - \Delta    
  \end{pmatrix}
  \begin{pmatrix}  \Psi_1(y) \\ \Psi_2(y)  \end{pmatrix} 
  =& E
    \begin{pmatrix}  \Psi_1(y) \\ \Psi_2(y)  \end{pmatrix} .
\label{eqeq2Delta}
\end{align} 
%
\begin{figure}[bt]
\centering
\includegraphics[width=0.4\textwidth]{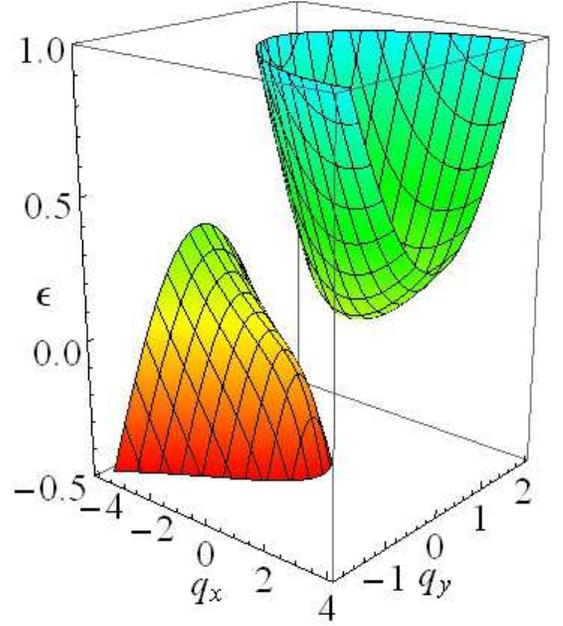} 
\caption{(Color online) 
Energy at $B=0$ as a function of $q_x$ and $q_y$ 
with parameters  $w_x =0.4$, $w_y=1$, $\alpha_{2}'=\alpha_{2}''=0.01$, $w_{0x}=-w_x$, $w_{0y}=0$,
and $\Delta=\pm0.3$. \\
}
\label{fig2qddelx}
\end{figure}
\begin{figure}[bt]
\vspace{0.5cm}
\centering
\includegraphics[width=0.42\textwidth]{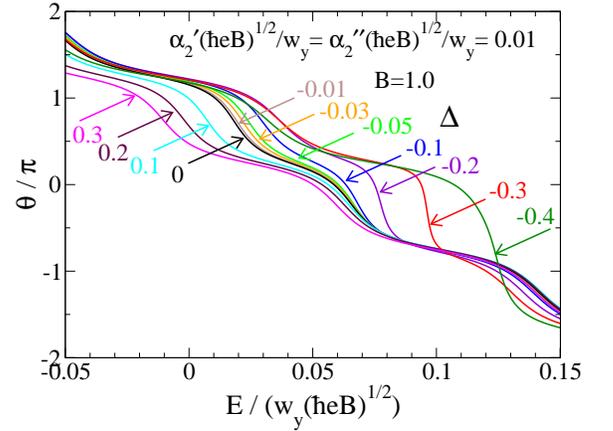} 
\caption{(Color online) 
Boundary condition 
$\theta$ at $y=y_{+}=20$ (Eq.~(\ref{eqbcy20})), which makes $|\Psi_{1,2} (y)| \to 0$ at $y \to -\infty$, 
as a function of the energy.
We take parameters  $w_x = w_y=1$, $\alpha_{2}'\sqrt{\hbar eB}/w_y=\alpha_{2}''\sqrt{\hbar eB}/w_y=0.01$, 
$\Delta=0.3$, $0.2$, $\cdots$, $-0.3$, $-0.4$, and $B=1$.
}
\label{figfig14}
\end{figure}
\begin{figure}[tb]
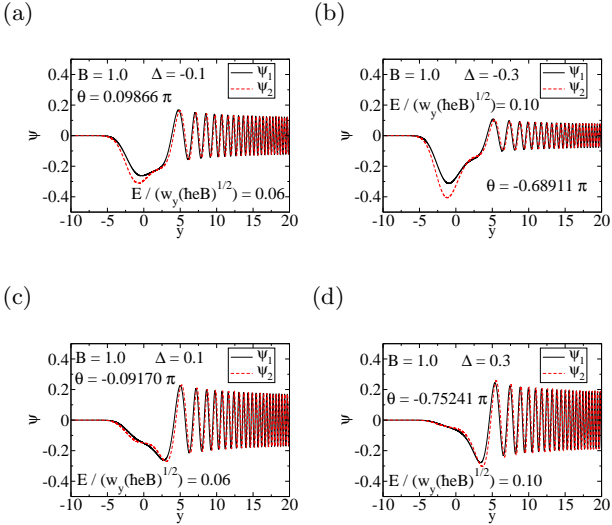

\centering
{\raggedright \ (a) \hspace{3.5cm} (b) \\ } \vspace{0.4cm}
\includegraphics[width=0.2\textwidth]{fig15anew.eps} \hfil 
\includegraphics[width=0.2\textwidth]{fig15bnew.eps} \\ \vspace{0.5cm}
{\raggedright \ (c) \hspace{3.5cm} (d) \\ } \vspace{0.4cm}
\includegraphics[width=0.2\textwidth]{fig15cnew.eps} \hfil 
\includegraphics[width=0.2\textwidth]{fig15dnew.eps}\\ 
\caption{(Color online) 
Wave functions for 
(a) $\Delta=-0.1$,  $E/(w_y\sqrt{\hbar eB})=0.06$,
(b) $\Delta=-0.3$, $E/(w_y\sqrt{\hbar eB})= 0.10$,
(c) $\Delta=0.1$,  $E/(w_y\sqrt{\hbar eB})=0.06$,
and
(d) $\Delta=0.3$, $E/(w_y\sqrt{\hbar eB})= 0.10$,
 Other parameters 
are  $w_x = w_y=1$, $\alpha_{2}'\sqrt{\hbar eB}/w_y=\alpha_{2}''\sqrt{\hbar eB}/w_y=0.01$, and
$B=1$. Boundary conditions at $y=20$ ($\theta$) are taken as in the case of $\Delta=0$ 
in section II B. 
 The wave functions have large amplitudes in $y \lesssim 0$ region, when $\Delta<0$ [(a) and (b)],
while no peaks are seen in  $y \lesssim 0$ region,  when $\Delta > 0$ [(c) and (d)]. 
}
\label{figfig15}
\end{figure}

Although the energy dispersion at $B=0$ does not depend on the sign of $\Delta$,
the quantized energies at $B \neq 0$ are not the same for $\pm \Delta \neq 0$.
We take $\alpha_2'=\alpha_2''=0.01$ and $B=1$ and we calculate the wave functions
numerically 
with the boundary condition at $y=20$,
as in Section \ref{sectionbc1}.
We plot the boundary condition $\theta$ to exist a non-divergent solution 
 as a function of energy in Fig.~\ref{figfig14}.
For $\Delta \lesssim -0.1$, $\theta$ changes in a narrow region of $E$,
which indicate that an almost-localized state exists at $y \lesssim 0$ as shown in Fig.~\ref{figfig15}~(a) and (b), 
while the
variation of $\theta$ as a function of $E$ becomes broad for $\Delta \geq -0.1$,
which indicate that an almost-localized state at $y \lesssim 0$
couples strongly to the continuous energy state at $y>0$ and an almost-localized state
ceases to exist at $y \lesssim 0$
 as shown in Fig.~\ref{figfig15}~(c) and (d).   
 We think that the 
 eigenstate with $n=0$ does not exist when $\Delta>0$,
but the almost-localized state exists at $y \lesssim 0$ when $\Delta \leq 0$.
 The effect of the tunneling would become important as $\Delta$ approaches to zero and 
the almost-localized $n=0$ state at $y \lesssim 0$ couples strongly to the continuous energy levels in $y>0$.
This situation that the $n=0$ mode exists only when $\Delta \leq 0$
 is similar to the model studied by Haldane\cite{Haldane1988},
where the zero mode exists either upper band or lower band depending on the sign of the 
mass, which is $\Delta$ in the present model, and the direction of the magnetic field.
In our model the nearly bound state with $n=0$ exists when $\Delta\leq 0$. 
The $n=0$ ($E=0$) state at $\Delta=0$
in Fig.~\ref{fige0}(a) is understood as the zero-mode of the almost-localized state at three-quarter
Dirac point, which couples strongly to the continuous states at $y \gtrsim 0$.
Note that the simultaneous changes of $B \leftrightarrow -B$, 
$y \leftrightarrow -y$, $\Psi_1 \leftrightarrow \Psi_2$, and
$\Delta \leftrightarrow -\Delta$ do not change Eq.~(\ref{eqeq2Delta}).

\section{Summary}
We study the quantized energy at the three-quarter Dirac point in the presence 
of external magnetic field $B$.
We obtain that the quantized energy is
proportional to 
$\alpha_2^{\frac{3}{5}}(nB)^{\frac{4}{5}}$ 
 (Eq.~(\ref{eqquantizedE})) by 
calculating the solution of the differential equation (Eq.~(\ref{eqeq2})) numerically.
We also obtain the approximate result  in the limit of $|\alpha_2 \sqrt{\hbar eB} /w_y| \ll 1$
as 
 $E \propto w_x^{-\frac{2}{5}} w_y^{\frac{4}{5}} \alpha_2^{\frac{3}{5}} (n \hbar eB)^{\frac{4}{5}}$
  (Eq.~(\ref{eqeq22x})), 
which is consistent with the result obtained 
in the previous paper\cite{Kishigi2017} by using the semiclassical quantization rule.
We show that the zero mode exists 
by studying the finite-gap system.
Since the three-quarter Dirac points with the finite gap appear as a pair when 
the time-reversal symmetry is not broken at $B=0$, sign of $\Delta$ is
positive at one  finite-gap three-quarter Dirac point and negative at another
point. As a result, there is one zero mode in the system when $B \neq 0$ and
$\Delta \neq 0$.

The quantization of the energy 
in the three-quarter Dirac point 
in a magnetic field
can be observed experimentally in quasi-two-dimensional organic superconductor
$\alpha$-(BEDT-TTF)$_2$I$_3$\cite{Kishigi2017} 
and ultra cold Fermi gas on a tunable optical lattice\cite{Tarruell2012}.

\appendix*
\section{another derivation of $E \propto  \alpha_2^{\frac{3}{5}} B^{\frac{4}{5}}$}
From Eq.~(\ref{eqeq2}) we formally obtain the equation
\begin{equation}
 \det \begin{pmatrix}
 M & N_{-}\\ N_{+} & M
 \end{pmatrix}
 =0,
\end{equation}
where
\begin{align}
 M &=  \frac{w_x}{w_y} y + \frac{\alpha_2' \sqrt{\hbar eB}}{w_y} y^2 - \frac{E}{w_y\sqrt{\hbar eB}}, \\
 N_{\pm} &=   \pm \frac{d}{dy} - \frac{w_x}{w_y} y + \frac{\alpha_2'' \sqrt{\hbar eB}}{w_y} y^2 
\end{align}
to get
\begin{align}
\frac{E}{w_y \sqrt{\hbar eB}} &= \frac{w_x}{w_y} y +  \frac{\alpha_2' \sqrt{\hbar eB}}{w_y} y^2
\nonumber \\
&\pm \sqrt{\left( \frac{-w_x}{w_y} y + \frac{\alpha_2''\sqrt{\hbar eB}}{w_y} y^2 \right)^2 - \frac{d^2}{dy^2}}
\end{align}
The almost localized state in $y<0$ is obtained by taking the expansion
\begin{align}
\left| \frac{d^2}{dy^2} \right| &\ll
\left( \frac{-w_x}{w_y} y + \frac{\alpha_2''\sqrt{\hbar eB}}{w_y} y^2 \right)^2.
\end{align}
Then we obtain 
\begin{equation}
 \frac{E}{w_y\sqrt{\hbar eB}} \sim \frac{\alpha_2 \sqrt{\hbar eB}}{w_y} y^2
 - \frac{w_y}{2 w_x} \frac{1}{(-y)} \frac{d^2}{dy^2},
 \label{eqapp}
\end{equation}
where we have used
\begin{equation}
\left| \frac{\alpha_2'' \sqrt{\hbar eB}}{w_y} y^2 \right|  
\ll \left| \frac{w_x}{w_y} y \right| .
\end{equation}
Taking a new variable $Y$ as
\begin{equation}
 y = \left( \frac{\alpha_2\sqrt{\hbar eB}}{w_y} \right)^{\sigma} \left( \frac{w_x}{w_y} \right)^{\nu} Y,
\end{equation}
and making the two terms in the right hand side 
of Eq.~(\ref{eqapp}) to be the same order in the dimensionless 
parameters $\alpha \sqrt{\hbar e B}/w_y$ and $w_x/w_y$, 
we obtain
\begin{equation}
	\sigma=- \frac{1}{5},
\end{equation}
and 
\begin{equation}
  \nu=-\frac{1}{5}.
\end{equation}
Then we obtain
\begin{align}
 E 
&\sim  w_y \sqrt{\hbar eB} \left( \frac{\alpha_2 \sqrt{\hbar eB}}{w_y} \right)^{\frac{3}{5}} 
\left( \frac{w_x}{w_y} \right)^{-\frac{2}{5}} \nonumber \\
&\times \left( Y^2 -\frac{1}{(-2Y)} \frac{d^2}{dY^2} \right).
\end{align}
Since $Y$ does not depend on any parameters, we obtain
\begin{equation}
  E \sim w_x^{-\frac{2}{5}} w_y^{\frac{4}{5}} \alpha_2^{\frac{3}{5}} (\hbar eB)^{\frac{4}{5}}.
\end{equation}
\bibliography{diracpoint}
\end{document}